\documentclass[aps,nofootinbib,groupaddress,floatfix,preprint]{revtex4-1}

\usepackage{graphicx}
\usepackage{amsmath}
\usepackage{dcolumn}
\usepackage{subfigure}
\usepackage{multirow}
\usepackage{mathtools}
\usepackage{epstopdf}
\usepackage{hyperref}
\usepackage{dcolumn, comment}% Align table columns on decimal point
\usepackage{bm}% bold math
\usepackage[left]{lineno}
\usepackage{mathptmx}

\usepackage{longtable}
\begin{document}
\title{Ultra-Low Dissipation Superfluid Micromechanical Resonator}

\author{F.~Souris}
\email{souris@ualberta.ca}
\affiliation{Department of Physics, University of Alberta, Edmonton, Alberta, Canada, T6G 2E9}

\author{X.~Rojas}
\altaffiliation[Present address: ]{Department of Physics, Royal Holloway University of London, Egham, Surrey TW20 0EX, UK}

\author{P.H.~Kim}
\affiliation{Department of Physics, University of Alberta, Edmonton, Alberta, Canada, T6G 2E9}

\author{J.P.~Davis}
\email{jdavis@ualberta.ca}
%\address{Department of Physics, University of Alberta, Edmonton, Alberta, Canada, T6G 2E9}
\affiliation{Department of Physics, University of Alberta, Edmonton, Alberta, Canada, T6G 2E9}

\begin{abstract}
Micro and nanomechanical resonators with ultra-low dissipation have great potential as useful quantum resources. The superfluid micromechanical resonators presented here possess several advantageous characteristics: straightforward thermalization, dissipationless flow, and \textit{in situ} tunability.  We identify and quantitatively model the various dissipation mechanisms in two resonators, one fabricated from borosilicate glass and one from single crystal quartz. As the resonators are cryogenically cooled into the superfluid state, the damping from thermal effects and from the normal fluid component are strongly suppressed. At our lowest temperatures, damping is limited solely by internal dissipation in the substrate materials, and reach quality factors up to 913,000 at 13 mK. By lifting this limitation through substrate material choice and resonator design, modelling suggests that the resonators should reach quality factors as high as 10$^8$ at 100 mK, putting this architecture in an ideal position to harness mechanical quantum effects.
\end{abstract}

%\pacs{67.25.bh, 67.60.gj, 67.60.an}

\maketitle
\linespread{1.2}

%\section{Introduction}

Recently there has been heightened interest in micro and nanomechanical systems as quantum resources, as opposed to traditional---classical---applications such as force \cite{Miao2012}, mass \cite{Chaste2012} or torque \cite{Kim2013} sensing. For example, mechanics in the quantum regime have demonstrated coupling to microwave qubits \cite{OConnell2010}, entanglement between phonons and photons \cite{Palomaki2013}, and quantum state transduction \cite{Lecocq2016}.  Long phonon lifetimes facilitate such quantum operations, hence resonators for quantum applications are generally made of low dissipation materials such as single crystal Si \cite{Meenehan2014} or stressed superconducting aluminum \cite{Cicak2010}.  Yet these materials are by no means ideal.  For example, two-photon adsorption in silicon is a significant limitation for optomechanics at low temperatures \cite{Meenehan2014}.  The search for improved materials with ultra-low mechanical dissipation is actively underway with studies of high-tension silicon nitride \cite{Verbridge2008} and diamond \cite{Mitchell2015,Burek2015}, for example.  Such materials should also have ultra-low dielectric loss in the telecom or microwave bands, to be compatible with cavity optomechanics \cite{Aspelmeyer2014}.  Development of new materials that meet these stringent requirements is therefore a key avenue to enabling further progress in quantum nanomechanics.   Recently it has been realized that a long-studied material, superfluid $^4$He, is actually one of the most promising candidates for mechanics in the quantum regime \cite{DeLorenzo2014,Rojas2015,Harris2016, Kashkanova2016,DeLorenzo2016}.

Below a critical temperature ($T_\lambda\simeq2.17$ K), liquid $^4$He undergoes a transition into a superfluid state, with macroscopic quantum coherence. This state of matter exhibits exotic properties such as frictionless flow below the so-called critical velocity, as well as extremely low mechanical dissipation at millikelvin temperatures.  These properties can be described using the two-fluid model, where the fluid is imagined to be composed of a conventional viscous fluid, the {normal} component $\rho_n$, and an inviscid fluid, the {superfluid} component $\rho_s$. At temperatures low compared to $T_\lambda$, where the normal fluid density vanishes ($\rho_n \rightarrow 0$), the mechanical properties of the superfluid are expected to be extraordinary. In recent groundbreaking work, mechanical quality factors reaching up to $Q=1.4\times10^8$ were demonstrated by De Lorenzo and Schwab \cite{DeLorenzo2014,DeLorenzo2016} in a superfluid acoustic resonator coupled to a superconducting microwave cavity, demonstrating the great potential of superfluid cavity optomechanics. 

While the work of De Lorenzo and Schwab \cite{DeLorenzo2014,DeLorenzo2016} is performed on gram-scale quantities of superfluid, applicable to the detection of high-frequency gravitational waves \cite{Singh2016}, quantum optomechanics experiments require larger zero-point fluctuations for measurement and control of quantum phenomenon: decreasing the mass of the mechanical system increases its usefulness as a quantum resource.  Furthermore, confinement to the nanoscale is enticing for study of superfluids themselves: enabling tests of finite-size scaling theory \cite{Gasparini2008} and proximity effects \cite{Perron2012} in superfluid $^4$He, and tests for new superfluid phases \cite{Vorontsov2007,Wiman2015} and surface Majorana fermions in $^3$He \cite{Chung2009,Wu2013,Mizushima2012,Levitin2013}.  

\begin{figure*}[t]
	\includegraphics[width=\linewidth]{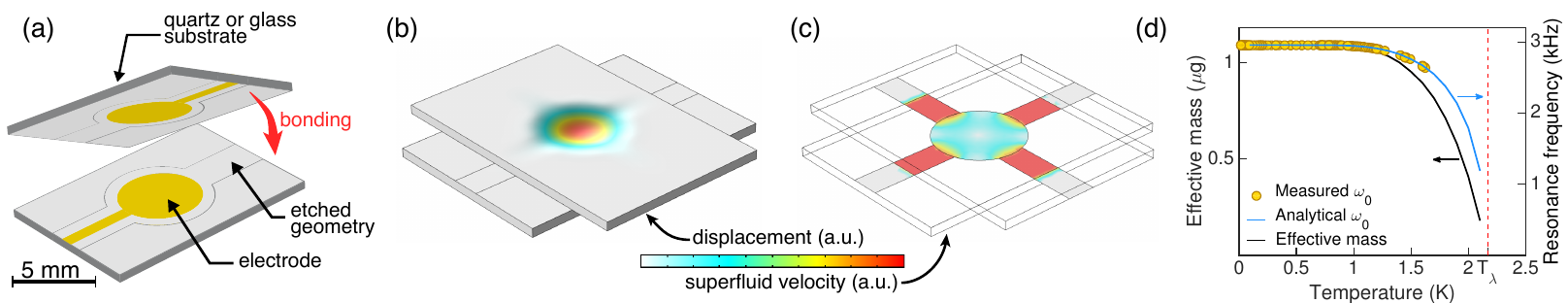}
	\caption{\linespread{0.8} \textbf{Helmholtz resonator characteristics} (a) Illustration of the resonator microfabrication scheme. The basin and the channels are etched into the substrate (borosilicate glass or single crystal quartz), chromium/gold electrodes are deposited and the two substrates are bonded together to create the basin for confining liquid helium. (b) Simulation showing the substrate deflection when a voltage is applied across the two electrodes. The electrostatic force bends the substrate inward, driving the Helmholtz resonance. (c) Simulation of the local superfluid velocity for the Helmholtz resonance. (d) Temperature dependence of the resonance frequency measured for the quartz device (yellow circles), along with the theoretical resonance frequency (blue curve).  The black curve is the corresponding effective mass of the Helmholtz mode, which grows with increasing superfluid density, $\rho_s/\rho$.}
	\label{fig:ResonanceMode}
\end{figure*}

Recent superfluid cavity optomechanics experiments have found that even the small dielectric constant of liquid helium is sufficient to study acoustic modes in fiber-cavities \cite{Kashkanova2016} and third sound in superfluid films \cite{Harris2016}---with significantly reduced effective masses as compared with Ref.~\cite{DeLorenzo2014}.  Yet neither of these experiments, nor our previous work \cite{Rojas2015}, achieved the ultra-low mechanical dissipation of De Lorenzo and Schwab \cite{DeLorenzo2014}. Therefore in reduced geometries, dissipation must be systematically revisited to determine the role of confinement. Here we show that a microgram effective mass superfluid Helmholtz resonator, in a slab geometry defined by microfabrication \cite{Duh2012,Rojas2014}, can achieve ultra-low mechanical dissipation, leading to quality factors up to $Q=9\times10^5$ at 13 mK.  For a $3$ kHz Helmholtz mode, this results in a mechanical dissipation rate of $\Gamma_m/2\pi = 3$ mHz---with a million times less effective mass than De Lorenzo and Schwab \cite{DeLorenzo2014,DeLorenzo2016}, see Table \ref{table1}---a phonon life time of $\tau=Q/\Omega_m\approx50$ s, and a thermal coherence time of $\tau_\textrm{th}=\hbar Q/k_\textrm{B} T_\textrm{bath} \approx530~\mu$s \cite{Aspelmeyer2014}.

{\renewcommand{\arraystretch}{1.2}
	\begin{table}[b]
		\begin{ruledtabular}
			\begin{tabular}{lcccc}
				\squeezetable
					Reference						& $\Omega_m/2\pi$	& Mass					& Volume				& $\Gamma_m/2\pi$\\
													& (kHz)				& (kg)					& (m$^3$)				& (Hz)\\
					\hline
					De Lorenzo \cite{DeLorenzo2014}	& 8.1				& $5.7\times10^{-3}$	& $4.0\times10^{-5}$ 	& 0.0006\\
					Harris \cite{Harris2016}		& $482$				& $2\times10^{-15}$		& $1.4\times10^{-17}$	& 106\\
					Kashkanova \cite{Kashkanova2016}& $317\times10^3$	& $3.8\times10^{-13}$	& $2.7\times10^{-15}$ 	& 4500\\
					De Lorenzo \cite{DeLorenzo2016}	& 8.1				& $5.7\times10^{-3}$	& $4.0\times10^{-5}$ 	& 0.00006\\
					This work						& 3					& $1.1\times10^{-9}$	& $7.3\times10^{-12}$	& 0.003\\
			\end{tabular}
		\end{ruledtabular}
		\caption{Comparison of superfluid mechanical resonators.}\label{table1}
	\end{table}
}

Furthermore, we are able to quantitatively model the sources of dissipation in this system and find that at millikelvin temperatures dissipation is dominated by two-level systems in the microfabricated substrate material.  With this knowledge it will be possible to engineer improvements that would allow this mechanical resonator---which is straightforward to thermalize to millikelvin temperatures and conceptually could be coupled to a microwave resonator---to achieve quality factors well above $10^7$, and thermal coherence times of tens of milliseconds, putting this architecture in an ideal position to harness mechanical quantum effects.

The premise of a Helmholtz resonator is that a confined fluid can act as a mass-spring system, with the spring constant given by a combination of the compressibility of the fluid and the containment vessel, and the mass given by the moving fluid in the channel. Helmholtz resonances are commonly experienced as the whistle produced when air is blown across the top of a bottle.  Confining superfluid $^4$He likewise leads to a Helmholtz mode, and allows the flexibility to engineer the mode by altering the geometry \cite{Rojas2015}.  Here the confinement is dictated by microfabricated borosilicate glass \cite{Duh2012,Rojas2014} or single-crystal quartz, with integrated drive electrodes \cite{Rojas2015}---see Fig.~\ref{fig:ResonanceMode}.  As we discuss in greater detail below, comparing devices fabricated from two different substrate materials allows us to discern the role of the substrate on dissipation of the Helmholtz resonator.

In the present experiment, a microfabricated basin of area, $A$, is etched into the substrate material and subsequently patterned with drive electrodes, which are spaced by a distance $h \approx 900$ nm apart after bonding, see Fig.~1 and Supplemental Material. The basin is connected to the surrounding helium reservoir by four channels, each having cross-sectional area, $a$, and length, $l$. These dimensions, together with the stiffness of the plate, $k_p$, and the relative helium stiffness, $\Sigma = (k_p/2)/k_{h}$, fully define the superfluid Helmholtz resonance frequency.  When a voltage is applied to the electrodes, the substrate bends inward due to the large electrostatic force. In the presence of superfluid helium, this deformation can be used to drive the Helmholtz mode, with an angular resonance frequency:
\begin{equation}\label{eq:HelmholtzResonanceFrequency}
		\Omega_m = \sqrt{\Big(\frac{4a}{l\rho}\Big)\,\frac{\rho_s}{\rho}\,\frac{k_p/2}{ A^2(1 + \Sigma) }}
\end{equation}
proportional to the superfluid density, $\rho_s/\rho$ \cite{Rojas2015}. That is, the normal fluid component, $\rho_n$, is viscously clamped due to the sub-micron confinement of the channels (see Fig.~1c and Supplemental Material), and therefore the resonance frequency increases as the temperature decreases and the superfluid fraction, $\rho_s/\rho$, grows \cite{Rojas2015}.  As $T\rightarrow0$ the effective mass of the Helmholtz mode also increases, reaching $1.1~\mu$g, as shown in Fig.~1d.

\begin{figure}[t]
		\includegraphics[width=0.5\textwidth]{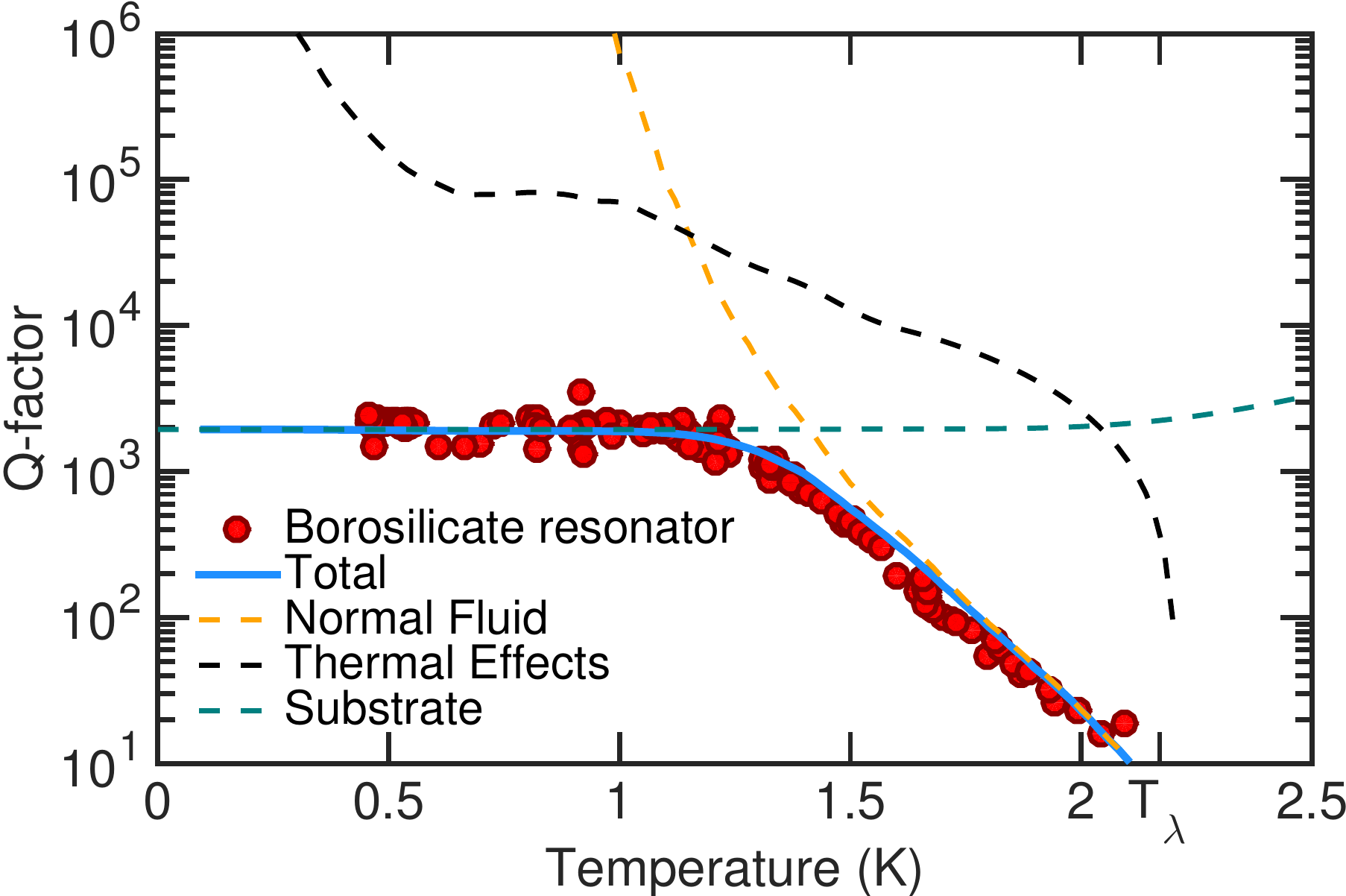}
		\caption{\linespread{1.0} The quality factor of the borosilicate resonator (red circles) was measured from 475 mK  to 2.09 K using a $^3$He fridge. The dashed orange curve shows the damping associated with the normal fluid residual motion and the dashed black curve is a model of thermal losses through the substrate, both described in the Supplemental Material. Because of the very small thermal conductivity of borosilicate, the normal fluid damping is the dominant loss mechanism. The dashed green curve is a prediction of the quality factor expected from internal dissipation in the borosilicate substrate.}\label{fig:DissipationGlass}
\end{figure}

\begin{figure}[b]
		\includegraphics[width=0.5\textwidth]{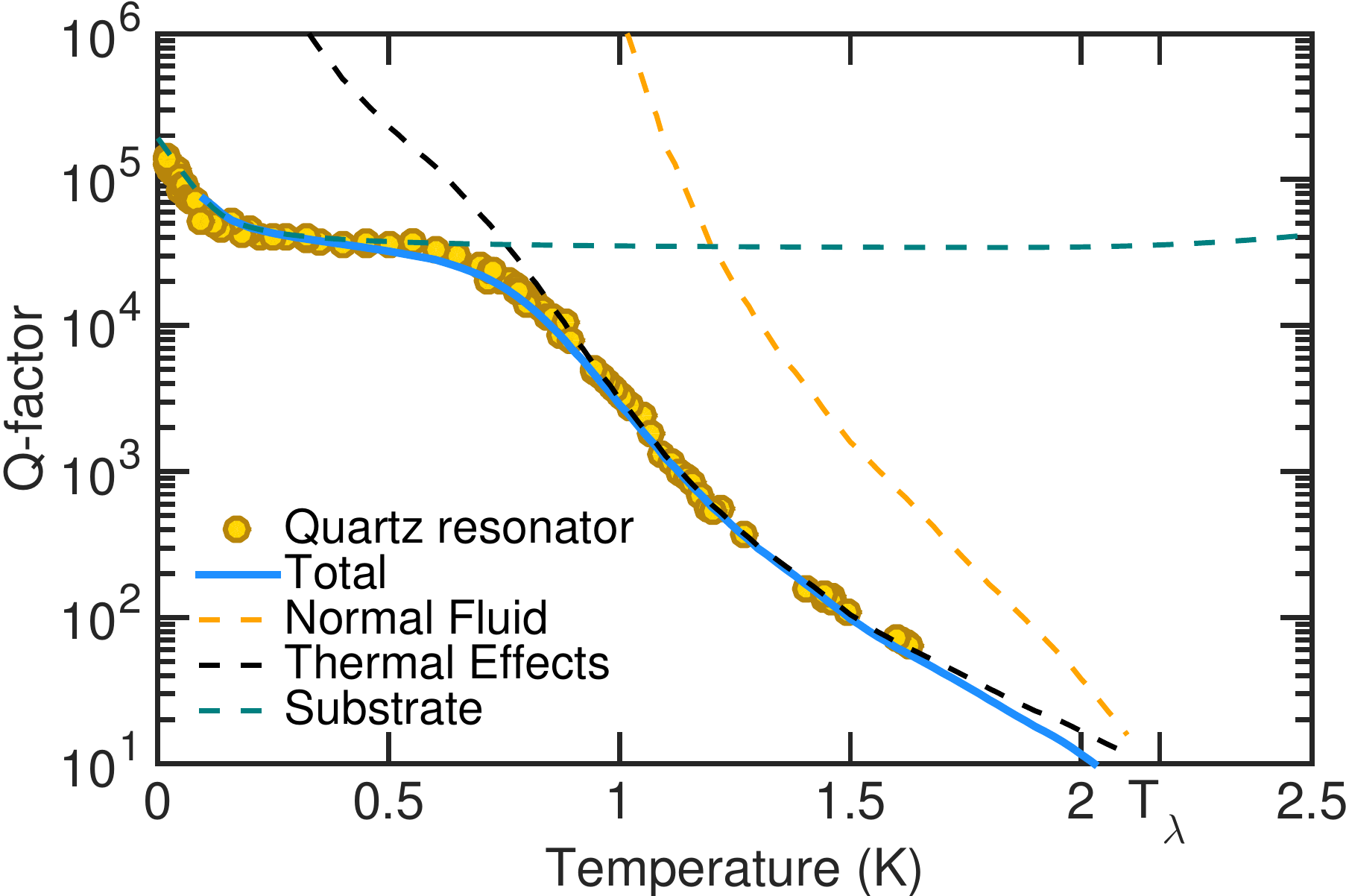}
		\caption{\linespread{1.0} An experiment similar to Fig. \ref{fig:DissipationGlass} is reproduced, using single crystal quartz substrates instead of borosilicate glass. The quality factor of the quartz resonator (yellow circles) is measured using a dilution refrigerator from 13 mK to 1.62 K. In contrast with the borosilicate resonator, the damping associated with the normal fluid residual motion (dashed orange curve) is now smaller than thermal losses through the substrate (dashed black curve). This is because while keeping the geometry and hence the normal fluid damping nearly constant, the thermal losses are increased because of the large thermal conductivity of crystalline quartz. The green dashed curve is a model of thermally excited two-level systems as described in the Supplemental Material. The behavior of the resonator is well reproduced over the entire experimental temperature range.}\label{fig:DissipationQuartz}
\end{figure}

Unlike the resonance frequency, the temperature dependence of the quality factor, $Q$, demonstrates complex behavior, as seen in Figs.~\ref{fig:DissipationGlass} and \ref{fig:DissipationQuartz}.  
For the first device studied here---fabricated from borosilicate glass---the quality factor increases monotonically until $\sim1.2$ K, and then plateaus. The initial rise in the $Q$ can be understood using the theory derived in Refs.~\cite{Brooks1979,Backhaus1997a}, which describes the dissipation introduced by the residual motion of the normal fluid in a superfluid Helmholtz resonance. Specifically, when the superfluid oscillates at $\Omega_m$ through the channel, the normal fluid is locked to the substrate if the viscous penetration depth, $\lambda = \sqrt{\eta/(\rho_n\Omega_m)}$, is larger than the channel height, $h/2$, with $\eta$ and $\rho_n$ being respectively the viscosity and the density of the normal fluid. As discussed in further details in the Supplemental Material, this is not exact and residual motion of the normal component does limit the quality factor to $Q_n = \frac{8\eta}{(h/2)^2} \frac{\rho_s}{\rho_n^2} \frac{1}{\Omega_m}$. Dissipation from normal fluid motion is represented in Fig.~\ref{fig:DissipationGlass} as an orange dashed line together with the experimental data. The model for $Q_n$ results in good agreement with the high temperature data, considering that no fit parameters were used.  As expected this dissipation mechanism vanishes at low temperature as $\rho_n/\rho\rightarrow0$. 

The second device---fabricated from crystalline quartz---also has a high temperature (700 mK to $T_{\lambda}$) quality factor that monotonically increases as the temperature is reduced, however its dissipation is dictated by a  different physical phenomenon, \emph{i.e.}~thermal losses. As the superfluid moves within the channel, with the normal fluid considered clamped, the amount of superfluid in the basin depletes and replenishes, producing temperature variations, $\Delta T$, due to the mechanocaloric effect. If the basin was perfectly thermally isolated from its environment, we would have $\Delta T =  \rho s T\Delta V / c_p$, with $s$ the specific entropy, $c_p$ the specific heat at constant pressure of the helium in the basin, and $\Delta V$ the amount of superfluid displaced. But the basin walls are not ideal thermal insulators and part of the heat produced leaks out of the basin into the superfluid reservoir, providing a thermal bath at the mixing chamber temperature, $T_{MC}$, and resulting in mechanical dissipation. This damping mechanism is maximized at high temperature, where $\Delta T$ is larger due the large specific entropy, $s$. Because the crystalline quartz substrate has a significantly larger thermal conductivity \cite{Gardner1981}, and is thinner than the borosilicate glass substrate, the heat propagates more easily between the basin and the surrounding superfluid, increasing this thermal loss channel. The model describing the quality factor $Q_{th}$ associated to thermal losses (see Supplemental Material), shown in Fig.~\ref{fig:DissipationQuartz} as a dashed black curve, agrees nicely with the quartz resonator data in the high temperature region. In this dissipation model, the only parameter adjusted is the area $A$ of the basin, scaled from the as-fabricated dimensions by a factor of two. This can reasonably be explained by the fact that as the heat propagates through the substrate of thickness $t$, the effective area contributing to thermal losses is increased. %A more detailed treatment, beyond the scope of this article, would require to solve the heat transport equation from the basin to the helium reservoir.

Remarkably, both thermal losses and normal fluid damping model predict exponential growth of the quality factor with lowering temperature that results in predicted $Q$'s at temperatures below $\sim400$ mK exceeding $10^6$. On the other hand, our device $Q$'s do not follow this exponential growth at low temperatures, and moreover the resonator fabricated from borosilicate glass saturates to $\approx1800$ at 1 K. It is well known that amorphous materials can have a strong acoustic attenuation at audio frequencies and low temperatures due to the presence of two-level systems interacting with the phonons \cite{Classen1994, Topp2014}, which is applicable to borosilicate glass.
By measuring a drum-like mechanical resonance mode of an evacuated device (at a significantly higher frequency of $\Omega_{{drum}}\simeq 220$ kHz), we have measured the internal dissipation $Q_{drum}^{-1}$ of the borosilicate material over a broad temperature range, from 450 mK to 100 K (see Supplemental Material). We found that the internal dissipation of the substrate is indeed dominated by the presence of two-level systems, in accordance with other studies \cite{Classen1994, Topp2014}, resulting in $Q_{drum}^{-1} = 7\times 10^{-4}$. This internal dissipation was found to be frequency independent in the temperature range studied, in agreement with the standard tunneling model \cite{Fefferman2008}, and should therefore be identical for the lower frequency Helmholtz resonance. As discussed in the supplementary information, since $1/(1+\Sigma) = 77 $\% of the potential energy is contained in the substrate stiffness for the borosilicate Helmholtz resonance,  two-level system damping accounts for $Q = Q_{drum}/0.77 = 1850$, explaining the saturation of $Q$ for the glass Helmholtz resonator. A more detailed model for the dissipation arising from two-level systems (see Supplemental Material) fits well with the low temperature data for the borosilicate device, as seen in Fig.~\ref{fig:DissipationGlass} as the green dashed curve.

Conveniently, the limitation to the quality factor of amorphous material can be lifted by employing a material for the substrate ideally exempt of any two-level systems. This possibility motivated the fabrication of the resonator using single crystal quartz as a substrate, since it has been shown to have low acoustic losses at cryogenic temperatures \cite{Galliou2011, Galliou2013}. Fig.~\ref{fig:DissipationQuartz} shows the temperature dependence of the quality factor for the quartz Helmholtz resonator studied from 13 mK to 1.7 K. As shown in Fig.~\ref{fig:DissipationQuartz} the behavior of the quartz resonator is very well described from 0.7 K to 1.7 K by a model of thermal losses through the substrate, with the quality factor increasing as the temperature decreases. The crystalline quartz substrate did substantially reduce the total dissipation, but as the resonator is cooled down the quality factor saturates below $\simeq700$~mK, to a value of $Q=4\times10^4$.

The plateau is followed by a further increase of the $Q$ below 100 mK. This increase at lower temperature rules out clamping loss as the mechanism responsible of the dissipation in the plateau region, as clamping loss are expected to be temperature independent since they depend on the stresses transmitted to the support structure. Instead we have found that this plateau region, and the subsequent rise in $Q$ at temperatures below 50 mK, can be well described using a model of two-level system induced dissipation in the substrate material. While we anticipated that such a dissipation mechanism should be negligible in single crystal quartz, two-level systems again seem to be the dominant source of dissipation below $\approx700$ mK. The dissipation in the crystalline quartz is modeled as a thermally-activated ensemble of two-level systems \cite{Tao2014}, with an energy splitting of $\Delta E = 1.3$ GHz, described in the Supplemental Material. In this model the dissipation occurs through an energy transfer between the two-level systems, through the modulation of $\Delta E$ by the oscillating elastic strain. As shown in Fig.~\ref{fig:DissipationQuartz}, this simple model describes the changes to the quality factor below 700 mK for the quartz, and combined with the thermal dissipation model, the behavior of the resonator can be  accurately accounted for over the entire experimental temperature range.

\begin{figure}[t]
		\includegraphics[width=0.7\linewidth]{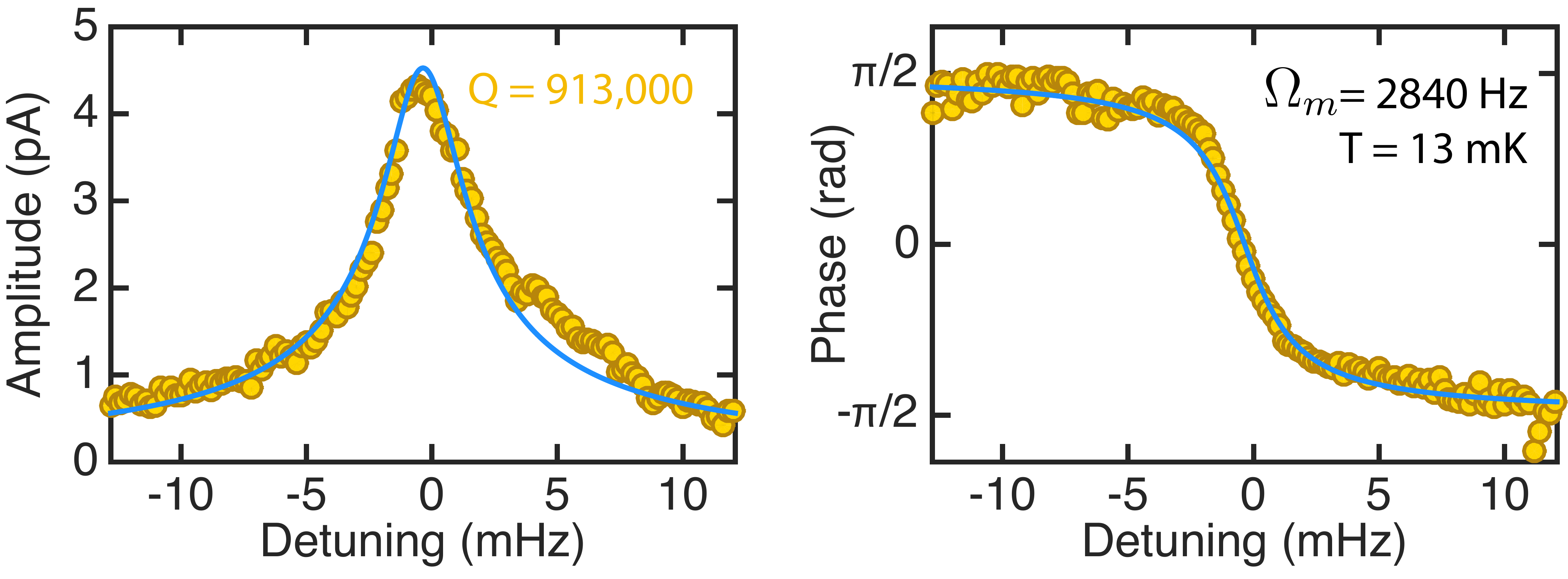}
		\caption{\linespread{1.0} Superfluid Helmholtz resonance detected at $T = 13$ mK, demonstrating a quality factor of Q = 913,000. For a resonance frequency $\Omega_m = 2840$ Hz, this results in a phonon lifetime longer than 50 seconds.}\label{fig:ResonanceRTheta}
\end{figure}

It is interesting to notice that for both the quartz and borosilicate resonators, the quality factor is ultimately limited by internal dissipation in the host material. Hence, reducing the amount of energy stored in the substrate should result in a drastic increase of the $Q$, up to $10^8$ at 100 mK in the case where the dissipation is only limited by thermal losses. This could be achieved by increasing the substrate stiffness through making the substrate thicker, reducing the diameter of the basin, or by patterning pillars that bridge between the two substrates in the area of maximal deflection. Reducing the substrate internal losses would also lead to an increase of the quality factor, for example by using diamond substrates that have been shown to have a $\Delta E$ of 13 GHz \cite{Tao2014}.  Likewise, in a subsequent experimental run, after warming up and cooling down the dilution refrigerator with the resonator left untouched in the sample cell, the quality factor of the same quartz Helmholtz was measured to be 913,000 at 13 mK, measured by a $^{60}$Co nuclear orientation thermometer, as shown in Fig.~\ref{fig:ResonanceRTheta}. This may be attributable to thermal cycling affecting the two-level systems in the substrate and therefore the quality factor.
 
In conclusion, we have fabricated and studied the behavior of two superfluid Helmholtz resonators with a sub-micron confinement. We show that their mechanical dissipation can be accurately modeled from $T_{\lambda}$, where the superfluid resonance takes place, down to $T = 13$ mK, while the mechanical dissipation spans over four orders of magnitudes. At low temperature, the damping is dominated by the internal dissipation of the substrate materials, and several solutions are offered to mitigate those effects for future experiments. This could result in quality factors as large as $10^8$ at 100 mK, to be compared with the 913,000 measured at 13 mK in the present experiment.

Such ultra-low dissipation fluid resonators, with the possibility to engineer confinement, could have numerous applications. For example, they could allow the creation of sensitive detectors to probe small volumes of superfluid $^3$He, predicted to undergo undiscovered phase transitions in confined geometries \cite{Vorontsov2007,Wiman2015}. Furthermore, the integration of this mechanical resonator into a microwave cavity, as in Ref.~\cite{Yuan2015}, would result in a superfluid optomechanical system with the zero point fluctuations of the acoustic field, $\Delta P_{zpf} =\sqrt{{\hbar\Omega_m}/{2V\chi}}$, being enhanced by the small volume of the resonator.  Because of the ultra-low dielectric loss in the telecom or microwave bands, such a superfluid optomechanical system could potentially be driven to extraordinary cavity enhanced cooperativities---key for all quantum measurement and control operations \cite{Aspelmeyer2014}.  Ultimately, ultra-low dissipation Helmholtz resonances could even be a potential astronomical tool, for the detection of continuous source of gravity waves, as reported by Singh \textit{et al.}~\cite{Singh2016}.

%\appendix

\section*{Supplement A: Resonators nanofabrication process and geometry}

\setcounter{equation}{0}

\renewcommand{\thefigure}{S\arabic{figure}}
\renewcommand{\theequation}{S\arabic{equation}}
\renewcommand{\thetable}{S\arabic{table}}
\renewcommand*{\citenumfont}[1]{S#1}
\renewcommand*{\bibnumfmt}[1]{[S#1]}

	\subsection{Definition of the geometry}
\begin{figure}[b]
			\includegraphics[width=\textwidth]{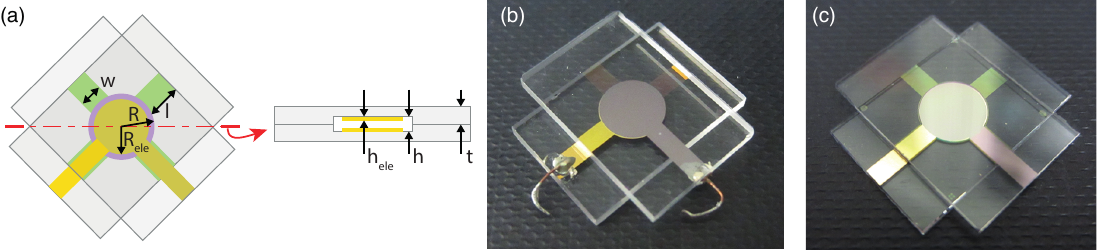}
			\caption{\linespread{1.0} The Helmholtz resonators are constructed by bonding two microfabricated substrates, forming a central basin and four channels.  Panel (a) defines the essential dimensions: the channels are depicted in green, the basin in purple, and the electrodes in yellow. These features are visible in the finished devices, shown in panels (b) and (c), for the borosilicate and the quartz resonators respectively. }\label{fig:PicturesResonators}
		\end{figure}

	The nanoresonators are built using standard nanofabrication techniques. The geometry is defined by optical lithography, etched into the wafer, and subsequently diced into chips that are bonded together to create the basin and channels of the Helmholtz resonator. The first resonator was fabricated from a $1.1$ mm thick borosilicate wafer and the second resonator from a $0.5$~mm thick single crystal Z-cut quartz wafer. In Fig.~\ref{fig:PicturesResonators} we define the relevant dimensions of the geometry and show a picture of each resonator. To facilitate comparison, the two resonators were designed to have nearly identical geometries. The two devices only differ by the wafer thickness, $t$, and the electrode radius, $R_{ele}$, given in Table \ref{tab:ResonatorGeometries} together with other relevant dimensions and properties. 
		
		{\renewcommand{\arraystretch}{1.2}
			\begin{table}[b]
				\begin{ruledtabular}
					\begin{tabular}{lcccc}
														& Symbol	& Units		& Quartz				& Borosilicate\\
					\hline
					Basin depth							& $h$ 		& (nm)		& 1014					& 1100\\
					Electrodes thickness 				& $h_{ele}$	& (nm)		& 50					& 100\\
					Basin radius 						& $R$		& (mm)		& 2.5					& 2.5\\
					Electrode radius 					& $R_{ele}$	& (mm)		& 2.3					& 2.4\\
					Channel width 						& $w$		& (mm)		& 1.6					& 1.6\\
					Channel length 						& $l$		& (mm)		& 2.5					& 2.5\\
					Channel cross sectional area 		& $a$		& (mm$^2$)	& $7.3 \times 10^{-4}$	& $7.2 \times 10^{-4}$\\
					Basin area 							& $A$		& (mm$^2$)	& 19.6					& 19.6\\
					Substrate thickness 				& $t$		& (mm)		& 0.5					& 1.1\\
					Substrate spring constant 			& $k_p$		& (N/m)		& $1.2 \times 10^{7}$	& $4.2 \times 10^{7}$\\
					\end{tabular}
				\end{ruledtabular}
				\caption{Summary of the dimensions and physical properties of the two resonators used.  Measurement of the substrate spring constant, $k_p$, is explained in Supplement B.}\label{tab:ResonatorGeometries}
			\end{table}
		}

	\subsection{Nanofabrication process}
\begin{figure}[t]
				\includegraphics[width=\textwidth]{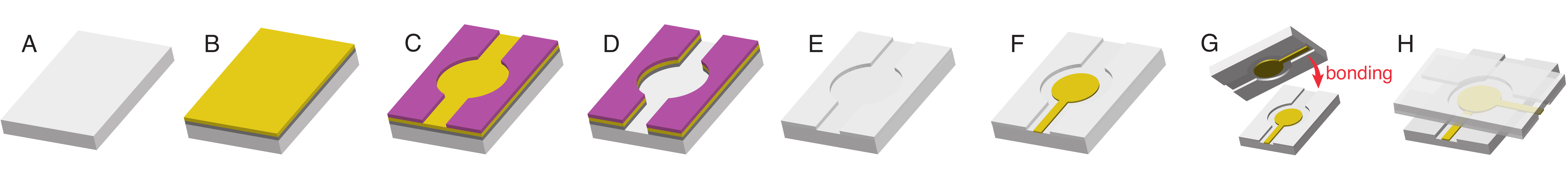}
				\caption{\linespread{1.0} Different steps of the nanofabrication process, as described in the text. White represents the wafer, yellow the metal masking layer and electrodes, and purple the photoresist.}\label{fig:NanofabricationProcess}
		\end{figure}

		The nanofabrication steps used to build the borosilicate and the quartz resonators are summarized in Fig.~\ref{fig:NanofabricationProcess}.
		The process starts with a cleaning of the bare quartz [borosilicate] wafer (step A) in a hot piranha solution (H$_2$SO$_4$ + H$_2$O$_2$). After cleaning, a chromium/gold masking layer is sputtered on the wafer (step B). Once developed, the positive photoresist HPR 504, shown in purple in Fig.~\ref{fig:NanofabricationProcess}, defines the geometry to be etched onto the wafer (step C).
		The exposed portion of the masking layer is then removed using chromium and gold etchants (step D), the resist is removed in acetone/IPA, and the wafer cleaned in a cold piranha solution. The basin and channels are then etched (step E) with a borofloat etchant [Silox Vapox III] having an etch rate of $133$~nm/min. The masking layer is then removed as in step D and the wafer cleaned again in a hot piranha solution prior to the electrodes deposition.
		A chromium/gold layer is deposited that will later create the electrodes. To shape the electrodes, a negative SU-8 photoresist is deposited, exposed and developed after careful alignment of the mask. The revealed part of the chromium/gold is etched away and the photoresist removed with a Remover PG solution. After a dip in a cold piranha bath, the final design is obtained (step F) and the wafer is ready for dicing and bonding. The bonding process is performed under a microscope by manually aligning and pressing two chips together (step G) to obtain the final resonator (step H).

\section*{Supplement B: Determination of the substrate stiffness}\label{sec:PlateConstantMeasurement}

	When a constant voltage bias, $V_{dc}$, is applied across the two electrodes of the resonator the substrate bends under the applied electrostatic load $q_{ele} = \epsilon_0/2\times\left(V_{dc}/(h-h_{ele})\right)^2$, where $\epsilon_0$ is the vacuum permittivity. Its mean deflection $\bar{x}$, as measured through the capacitance change between the two electrodes, can be used to deduce the substrate spring constant $k_p$. 
	The mean deflection $\bar{x} = \pi R^2 q_{ele}/k_p$ of the substrate is obtained by integrating the deflection of a doubly clamped loaded circular plate \cite{Roarks1992}, and since $\bar{x} \ll (h-h_{ele})$, the capacitance $C$ of the resonator with a bent substrate is approximated by:
	\begin{align}\label{eq:CapacitanceBias}
		C 	&\simeq \frac{\epsilon_0 \pi R_{ele}^2}{(h-h_{ele})}\Big( 1+ \frac{2\bar{x}}{(h-h_{ele})}\Big),\\
			&\simeq C_0 + \gamma\,V_{dc}^2.
	\end{align}
\noindent 
$C_0=\epsilon_0 \pi R_{ele}^2/(h-h_{ele})$ is the capacitance of an unbiased resonator and $\gamma = (C_0/(h-h_{ele}))^2\times(R/R_{ele})^2/k_p$ the capacitance change as a function of $V_{dc}$.
	The deflection profile is calculated by considering a uniform load distributed over the entire basin area, instead of applied only to the electrodes. Since the difference in the load is located at clamped edges, the error in the deflection profile is expected to be negligible.

\begin{figure}[t]
		\includegraphics[width=0.8\textwidth]{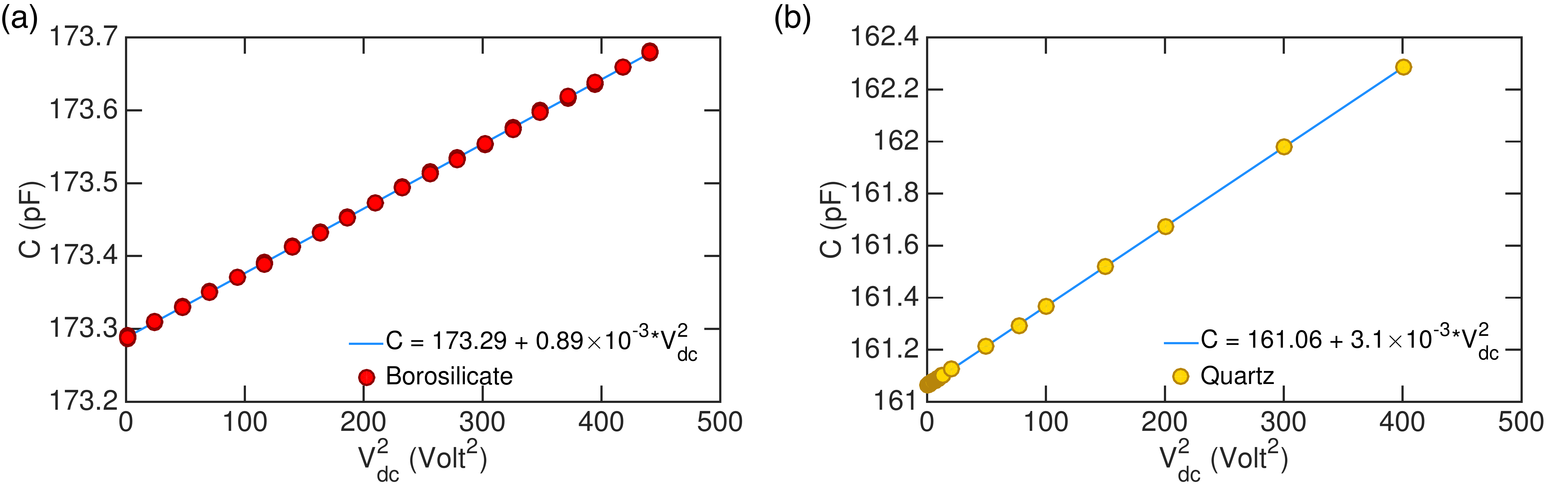}
		\caption{\linespread{1.0} To determine the substrate stiffness, we measure the device capacitance $C$ as a function of the square of the bias voltage: $V_{\textrm{dc}}^2$. The blue line is a linear fit to the data, from which the slope, $\gamma$, is extracted and used to calculate the substrate stiffness, $k_p$. We find $k_p = 4.2\times10^7$ N/m for the borosilicate device and $1.2\times10^7$ N/m for the quartz.}\label{fig:BiasMeasurement}
	\end{figure}

	The measurements of $C$ as a function of $V_{dc}$, presented in Fig.~\ref{fig:BiasMeasurement}, are performed at 4 K and are valid down to mK temperature within 0.1\% uncertainty, since the elastic properties of borosilicate and quartz depend very weakly on $T$ in this temperature range \cite{Raychaudhuri1984}. Also since the voltage used for the capacitance measurement is much smaller than $V_{dc}$, it is neglected in Eq.~\ref{eq:CapacitanceBias}.
	
	The capacitance shows a linear behavior as a function of $V_{dc}^2$, as expected for $\bar{x} \ll (h-h_{ele})$. The slope $\gamma$ is used to deduce the substrate spring constant $k_p = C_0^2\times(R/R_{ele})^2/[\gamma (h-h_{ele})^2]$, with the factor $(R/R_{ele})^2$ arising from the fact that the electrode does not cover the entire basin area. From these measurements we obtained $k_p = 1.2 \times 10^{7}$~N/m for the quartz substrate and $4.2 \times 10^{7}$~N/m for the borosilicate, as summarized in Table \ref{tab:ResonatorGeometries}.

\section*{Supplement C: Dynamics of the superfluid resonator}
	\subsection{Mass, stiffness and resonance frequency of the resonator}
		A theoretical model describing in detail the reactive and dissipative behavior of a superfluid Helmholtz resonator is derived in Refs.~\cite{Backhaus1997,Backhaus1997b}, including the effects of the normal fluid, thermal expansion, and compressibility.  As mentioned by the authors, the final result of those calculations is quite cumbersome.  We simplify their result by making few assumptions in application of this model to our resonators. For the current geometry, the main assumptions are that the volume of the superfluid reservoir, $V_{res}$, surrounding the resonator is large compared to the volume of the basin, \emph{i.e.} $V\ll V_{res}$, and that we can---at first---neglect dissipative effects to describe the dynamics of the system.

		From a simplified point of view, the resonator can be regarded as a mass on a spring system, with the mass being the superfluid moving in the channels (considered as incompressible within the channels) such that $m = 4\,la\rho_s$. The spring constant, $K$, with respect to the superfluid displacement, $y$, arises from a combination in series of the substrate stiffness, $K_p$, plus the effect of the compressibility of the fluid inside the basin given by $K_h$.
		The contribution of the two substrates to the stiffness is given by:
		\begin{equation}
			K_p = \Big(\frac{\rho_s}{\rho}\Big)^2 \, \frac{(4a)^2}{A^2} \, \frac{k_p}{2},
		\end{equation}
		where $k_p$ is the substrate stiffness with respect to the mean deflection $\bar{x}$ as defined earlier in Supplement B. In a similar way, we define the contribution of the fluid compressibility:
		\begin{equation}
				K_h = \Big(\frac{\rho_s}{\rho}\Big)^2 \, \frac{(4a)^2}{A^2} \, k_h,
		\end{equation}
		with $k_h= \frac{A^2}{V\chi}$ the stiffness with respect to $\bar{x}$. The resulting stiffness is:
		\begin{equation}
			K = \frac{(4a)^2}{A^2}\,\Big(\frac{\rho_s}{\rho}\Big)^2 \, \frac{k_p/2}{ 1 + \Sigma },
		\end{equation}
			where $\Sigma = (k_p/2)/k_h$ represents the repartitioning of the potential energy between the substrate and the compressed fluid. Therefore the Helmholtz angular resonance frequency is given by:
			\begin{equation}\label{eq:HelmholtzResonanceFrequency}
			\Omega_m = \sqrt{\frac{K}{m}} =\sqrt{\Big(\frac{4a}{l\rho}\Big)\,\frac{\rho_s}{\rho}\,\frac{k_p/2}{ A^2(1 + \Sigma) }}.
		\end{equation}

	\subsection{Adjustment of the theoretical model to the measured resonance frequency}
		One particularity of the superfluid Helmholtz resonator, compared to other low temperature mechanical resonators, is that its resonance frequency $\Omega_m$ can be adjusted \textit{in situ} by changing the pressure or the temperature of the superfluid. Indeed, temperature and pressure can strongly affect several thermodynamic functions defining to the Helmholtz resonance frequency, such as the superfluid fraction $\rho_s/\rho$, the density $\rho$, and the compressibility $\chi$. In the following description, and if not mentioned explicitly in the text, all the thermodynamic properties for liquid helium are adapted from \cite{Brooks1977} and interpolated at the corresponding values of $T$ and $P$. 

\begin{figure}[t]
		  	\includegraphics[width=0.8\textwidth]{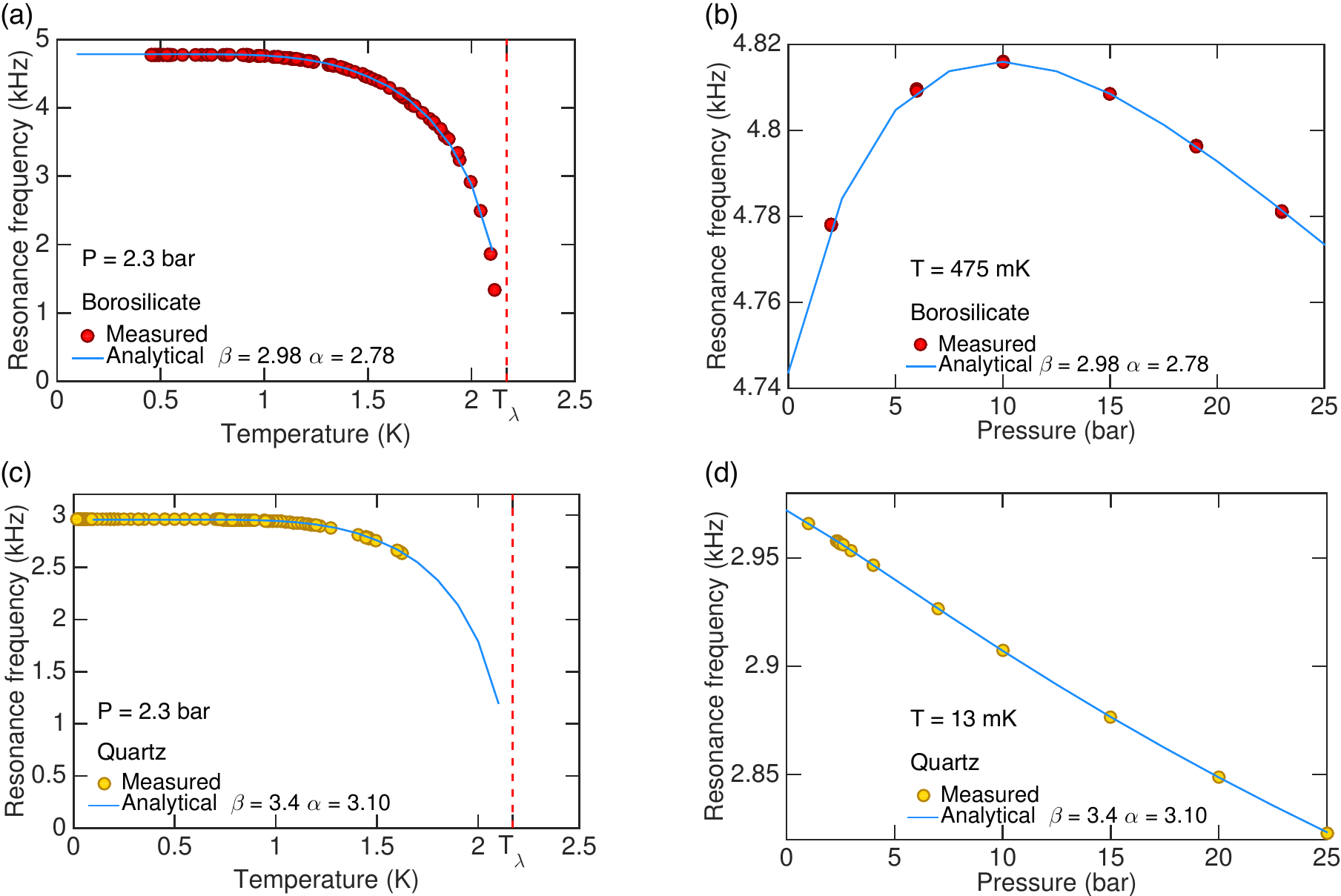}
			\caption{\linespread{1.0} For the borosilicate (a,b) and quartz (c,d) resonators the resonance frequency was measured as a function of pressure and temperature. (a) BorosilicateBorosilicate resonator frequency versus temperature with the pressure fixed at $P = 2.3$ bar. (b) Borosilicate resonator frequency versus pressure with the temperature at $T = 475$ mK. (c) Quartz resonator frequency versus temperature with the pressure set at $P = 2.3$ bar. (d) Quartz resonator versus pressure with the temperature set at $T = 13$ mK.  The blue curves are fits to Eq.~\ref{eq:HelmholtzResonanceFrequencyFit} with two adjustable parameters for each device.  Excellent agreement is found for both resonators for a wide pressure and temperature range. For the borosilicate resonator we find $\beta = 2.98$ and $\alpha = 2.78$, and for the quartz resonator $\beta = 3.4$ and $\alpha = 3.1$. }\label{fig:FitFrequency}
		\end{figure}

		To verify how Eq.~\ref{eq:HelmholtzResonanceFrequency} applies to a Helmholtz resonator with a slab geometry, with a basin volume much smaller that the one used in previous work \cite{Avenel1985, Brooks1979, Beecken1987}, we measured $\Omega_m$ for different $P$, $T$ parameters and adjusted Eq.~\ref{eq:HelmholtzResonanceFrequency} to this entire dataset. More precisely, for each resonator, we performed a temperature scan at constant $P$ and a pressure scan at constant $T$, as shown in Fig.~\ref{fig:FitFrequency}.
		Eq.~\ref{eq:HelmholtzResonanceFrequency} is fit to our measurements by introducing two parameters $\alpha$ and $\beta$, corresponding respectively to a scaling of $\Omega_m$ and a correction to the parameter $\Sigma$, such that:
		 \begin{equation}\label{eq:HelmholtzResonanceFrequencyFit}
			\Omega_m =\sqrt{\alpha \Big(\frac{4a}{l\rho}\Big)\,\frac{\rho_s}{\rho}\,\frac{k_p/2}{ A^2(1 + \beta \Sigma) }}.
		\end{equation}

		The fits shown in Fig.~\ref{fig:FitFrequency} have a good agreement with the data and Eq.~\ref{eq:HelmholtzResonanceFrequencyFit} seems to adequately describe the behavior of $\Omega_m$. For the borosilicate resonator, we obtain $\beta = 2.98$, $\alpha = 2.78$ and for the quartz $\beta = 3.4$, $\alpha = 3.1$. It is worth noticing that in both cases, the correction parameters $\alpha$ and $\beta$ are much larger than one, but are similar between devices. In our geometry, the basin volume is relatively close to the channel volume, and the channel length is of the order of the basin diameter. In this situation, end corrections may be necessary to properly describe the flow.  We therefore attribute the nonunitary values of $\alpha$ and $\beta$ to the lack of end corrections in our model.

		The corrected parameter $\beta\Sigma = \beta(k_p/2)/k_{h}$ gives the ratio of the substrate spring constant to the helium spring constant, which determines the percentage, $1/(1+\beta\Sigma)$, of potential energy stored in the bending of the substrate. In the case of the quartz resonator, at $2.3$~bar about $91$\% of the total potential energy is stored in the substrate, compared to $77$\% for the borosilicate.

\section*{Supplement D: Dissipative effects in the superfluid resonator}
	\subsection{Normal fluid dissipation}
			In Supplement C the Helmholtz resonance frequency, $\Omega_m$, was derived assuming that the normal component of the fluid is clamped to the substrate and does not contribute to the motion. This condition is fulfilled when the channel height, $h_c$, is large compared to the viscous penetration depth, $\lambda = \sqrt{\eta/(\rho_n\Omega_m)}$, where $\eta$ is the viscosity of the normal fluid. $\rho_n$ and $\Omega_m$ depend strongly on temperature, especially near $T_\lambda$ and it is not straightforward to \textit{a priori} predict the behavior of the viscous penetration depth in this temperature domain.
			\begin{figure}[b]
					\includegraphics[width=0.4\textwidth]{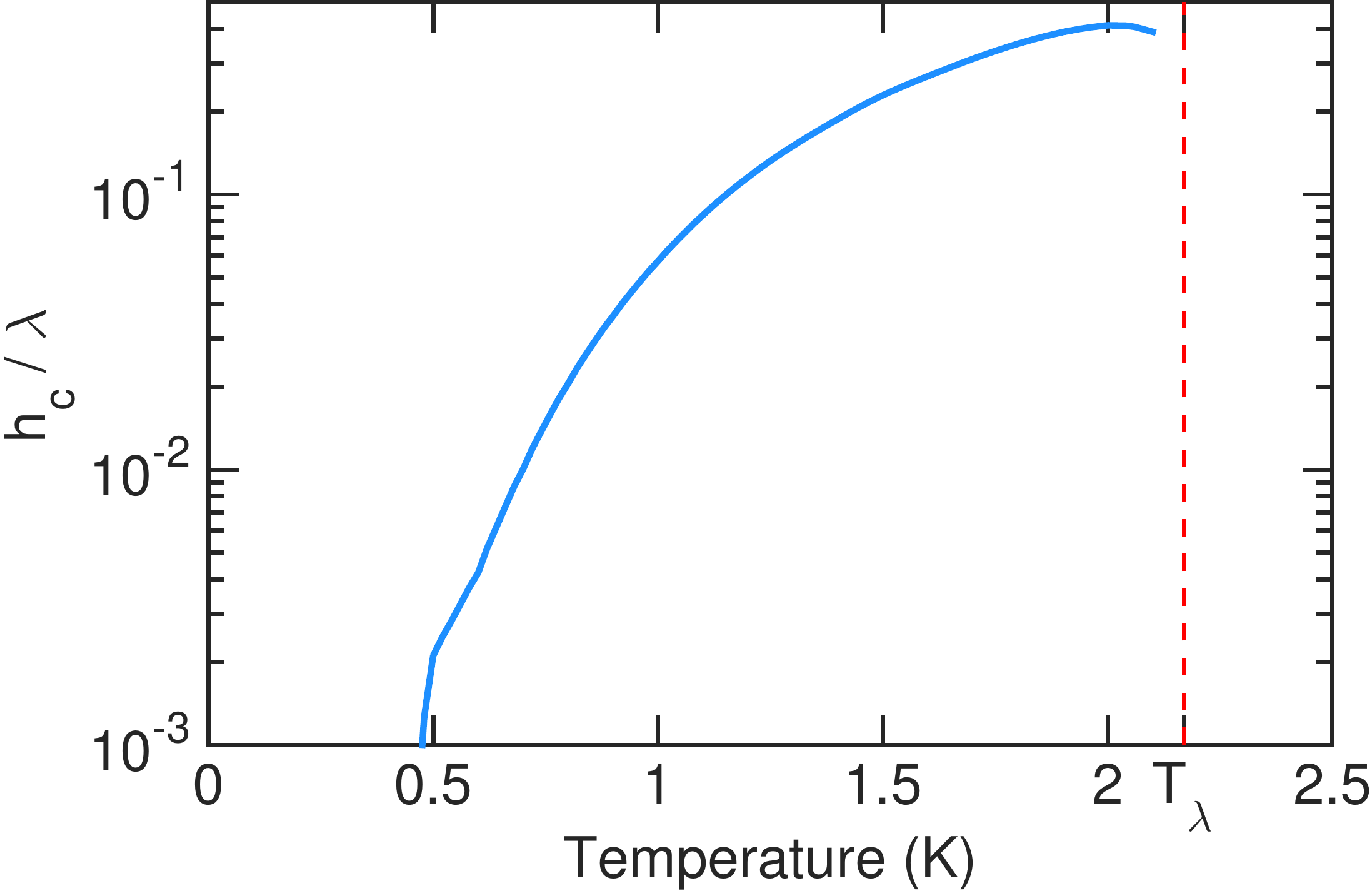}
					\caption{\linespread{1.0} When the channel height $h_c$ is small compared to the viscous penetration depth $\lambda$, the normal fluid is clamped by its own viscosity. This figure shows the ratio $h_c/\lambda$ as a function of temperature. Below 0.5 K, the normal fluid is completely locked by the diverging value of $\lambda$.}\label{fig:ViscousPenetration}
			\end{figure}
			Therefore in Fig.~\ref{fig:ViscousPenetration} we calculate the ratio of the channel height, $h_c$, to the viscous penetration depth, $\lambda$, as a function of $T$. Below $0.5$~K $\lambda$ diverges, since the normal fluid fraction vanishes, and it is reasonable to assume that the normal fluid is fully clamped. However, above 0.5 K, the viscous penetration depth is reduced and at $T_\lambda$ it is only half of $h_c$. Experimentally, as shown above, the behavior of the resonator frequency is well modeled when considering that only the superfluid is moving through the channels, and no corrections from the normal fluid are required. However, the normal fluid not only affects the resonator frequency, it also damps the resonator motion.

			When the residual normal fluid motion and the normal fluid fraction ($\rho_n/\rho$) are sufficiently large, viscous damping can become a non-negligible source of dissipation. As mentioned in Ref.~\cite{Brooks1977}, in the low damping limit the quality factor can be estimated by comparing the average rate of energy dissipation, $\langle D_n \rangle$, due to normal-fluid damping to the total amount of energy stored in the resonator, $E$, such that $Q_n = E\Omega_m/\langle D_n \rangle$. By applying the theory of Ref.~\cite{Brooks1977} to our resonator geometry, $Q_n$ can be written as:
			\begin{equation}\label{eq:Qnormal}
					Q_n = \frac{8\eta}{(h/2)^2} \frac{\rho_s}{\rho_n^2} \frac{1}{\Omega_m} \left[ 1 + \frac{\rho_s}{\rho_n}\frac{\alpha_P sT}{c_P} - \frac{\rho_s}{\rho_n} \rho\frac{A^2(1+\Sigma)}{V k_p/2} \frac{s^2T}{c_P} \right]^{-1},
			\end{equation}
			with $\alpha_P$ the isobaric thermal expansion coefficient of liquid helium, $c_P$ the specific heat per unit mass at constant pressure, and $s$ the the specific entropy.
			For our experimental conditions, the two last terms in Eq.~\ref{eq:Qnormal} can be neglected and $Q_n$ therefore reduced to:
			\begin{equation}\label{eq:QnormalSimplified}
				Q_n = \frac{8\eta}{(h/2)^2} \frac{\rho_s}{\rho_n^2} \frac{1}{\Omega_m}.
			\end{equation}
			Values for $\eta$ are extracted from Ref.~\cite{Donnelly1998}, and values for $\Omega_m$ are obtained from the fitting described in supplement C. Note that the viscous damping is directly proportional to the normal fluid density and consequently vanishes at low temperatures.

\subsection{Thermal effects in a Helmholtz resonator}
	When the superfluid oscillates within the channel, the kinetic energy of the superfluid in the channel is converted into potential energy that is stored in the substrate and the superfluid. Ideally, all the potential energy is released and does not lead to any dissipative effects. However, when the superfluid moves within the channel---with the normal fluid locked to the substrate---it drives a temperature difference, $\Delta T$, between the basin and the helium reservoir due to the resulting imbalance in superfluid ratios: an effect known as the mechanocaloric effect. Because of this temperature difference, heat flows from the basin to the reservoir and is a source of loss. We label $Q_{th}$ the quality factor related to this phenomenon. Calculation of $Q_{th}$ is based on Ref.~\cite{Backhaus1997}, in which a careful theoretical analysis of the response of a Helmholtz resonator is carried out. %The results presented here are limited to the case where thermal effects are the only source of dissipation and the normal fluid is considered fully clamped to the substrate.

	The temperature difference, $\Delta T$, has a reactive component in the form of a fountain pressure that acts as an additional spring constant, altering the resonator stiffness and resonance frequency. The importance of this fountain effect compared with the substrate spring constant is evaluated through the parameter $\Gamma_{th}$:
	\begin{equation}
		\Gamma_{th}^2 = \frac{\rho^2 s^2 T A^2 (1+\Sigma)}{C_{th}\,k_p/2},
	\end{equation}
	which is negligible for our resonator parameters, in that it does not alter $\Omega_m$.
	The time constant over which this temperature difference returns to equilibrium is given by $\tau_{th}=R_{th}C_{th}$, where $R_{th}$ is the total thermal resistance between the basin and the reservoir and $C_{th}$ is the total heat capacity of the basin. If the driving frequency $\Omega_m$ is small compared to $1/\tau_{th}$, the heat loss over a cycle becomes important leading to energy dissipation. This is described by the parameter $\Phi_{th} = 1/(\Omega_m \tau_{th})$.

	A simplified expression for $Q_{th}$ based on $\Phi_{th}$ and $\Gamma_{th}$ can be obtained in two different limits: in the low temperature limit, the specific entropy is small leading to $\Gamma_{th}^2\ll1$; in the high temperature limit, entropic effects dominate and we use instead $1 + \Gamma_{th}^2 \gg \Phi_{th}^2$. Therefore we have:
	\begin{equation}
		Q_{th} =
		\begin{cases}
			\dfrac{1+\Phi_{th}^2}{\Phi_{th}\Gamma_{th}^2}\,\left[ 1 + \dfrac{\Gamma_{th}^2}{2(1+\Phi_{th}^2)}\right]& \text{if $\Gamma_{th}^2\ll1$},\\
			\\
			\dfrac{(1 + \Gamma_{th}^2)^{3/2}}{\Phi_{th}\Gamma_{th}^2}\,\left[ 1 - \dfrac{(\Gamma_{th}^2 - 2)}{4(1+\Gamma_{th}^2)}\left(\dfrac{\Phi_{th}^2}{1+\Gamma_{th}^2}\right)\right]& \text{if $1 + \Gamma_{th}^2 \gg \Phi_{th}^2$}.
		\end{cases}
	\end{equation}
	In our case, since $\Gamma_{th}^2$ never exceeds 0.2 (even close to $T_\lambda$ where it is at its maximum), the low temperature approximation holds over the entire range of temperature and is therefore used to describe the thermal losses in our resonator.

\begin{figure}[b]
			\includegraphics[width=0.3\textwidth]{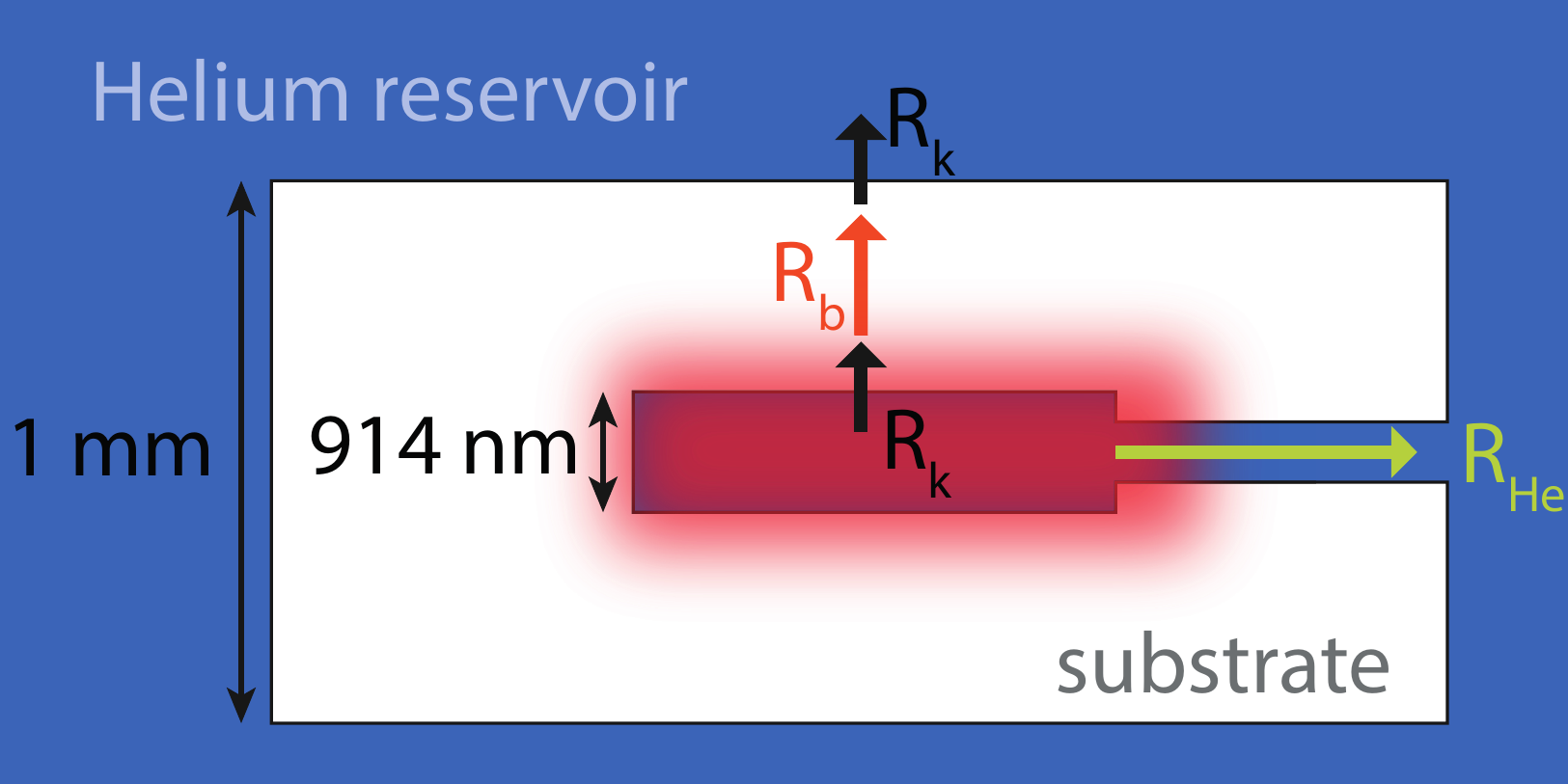}
			\caption{\linespread{1.0} The oscillation of the superfluid in the channel produces temperature changes inside the basin. The heat can flow in or out the basin by following two main paths, as shown in this sketch of the device. Heat propagation through the substrate requires overcoming two boundary resistances, $R_k$, and the bulk resistance, $R_b$. Heat flow through the channels is dictated by the thermal resistance of helium, $R_{He}$. }\label{fig:ThermalLosses}
	\end{figure}

	The calculation of $Q_{th}$ also requires knowing the total thermal resistance $R_{th}$ between the basin of the resonator and the reservoir surrounding it. $R_{th}$ depends on the geometry of the resonator, and the thermal conductivities of the substrate and liquid helium. Considering the geometry of the resonator, we can identify two paths through which the heat can flow from the basin to the reservoir, as illustrated in Fig.~\ref{fig:ThermalLosses}.
	
	The first path propagates through the superfluid/substrate interface, with Kapitza resistance $R_k$, the bulk substrate, with thermal resistivity $R_b$, and the substrate/superfluid interface, again with Kapitza resistance $R_k$. Since those resistances are in series, the total resistance of the first path is given by $R_{th_1}=2R_k + R_b$.
	The second path, illustrated in Fig.~\ref{fig:ThermalLosses}, propagates through the superfluid in the channel, with thermal resistivity $R_{He}$. The total thermal resistance of the second path is therefore simply $R_{th_2}=R_{He}$.
	Since each resonator has in total two substrates and four channels, the total thermal resistance, $R_{th}$, is:
	\begin{equation}\label{eq:TotalThermalResistance}
		{R_{th}} = \left(\frac{2}{R_{th_1}} + \frac{4}{R_{th_2}}\right)^{-1} = \left(\frac{2}{\left(2R_k + R_b\right)} + \frac{4}{R_{He}}\right)^{-1}.
	\end{equation}

	{\renewcommand{\arraystretch}{1.2}
		\begin{table}[b]
			\begin{ruledtabular}
				\begin{tabular}{lclccc}
				&	Quantity	& 	Units	& Quartz\footnote{The Kapitza resistivity is extracted from Ref.~\cite{Pollack1969} and the bulk thermal conductivity from Refs.~\cite{Hofacker1981,Gardner1981}.}						& Borosilicate\footnote{The Kapitza resistivity for borosilicate is taken to be the same as quartz and the bulk thermal conductivity is obtained from Ref.~\cite{Pobell2007}.}				& Helium\footnote{The thermal conductivity of helium in small channels is extracted from Ref.~\cite{Greywall1981}.}\\
				\hline
				Kapitza resistivity with $^4$He 		& $r_k$		& (cm$^2\cdot$K$\cdot$W$^{-1}$) 		& $17.5\,T^{-3.6}$			& $17.5\,T^{-3.6}$					& -\\
				Kapitza resistance with $^4$He 			& $R_k$		& (K$\cdot$W$^{-1}$)				& $89.1\,T^{-3.6}$			& $89.1\,T^{-3.6}$			& -\\
				Bulk thermal conductivity 				& $\kappa$	& (W$\cdot$cm$^{-1}$$\cdot$K$^{-1}$)	& $0.12\,T^{2.7}$			& $0.25 \times 10^{-3}T^{1.91}$	& $20\,h_c\,T^3$\\
				Bulk thermal resistance 				& $R_b$		& (K$\cdot$W$^{-1}$)				& $2.39\,T^{-2.7}$			& $2.05 \times 10^{3}T^{-1.91}$		& $3.12 \times 10^{7}T^{-3}$

				\end{tabular}
			\end{ruledtabular}
			\caption{\linespread{1.0} Summary of the thermal resistances encountered by the heat flowing from the basin to the surrounding helium reservoir. The thermal conductivity of helium in small channels depends linearly of the channel diameter \cite{Greywall1981}, since it is limited by the mean free path of phonons. Here we use $h_c$ as the channel diameter.}\label{tab:TabThermalConductivities}
		\end{table}
	}

	The bulk thermal properties and boundary (Kapitza) resistance for quartz, borosilicate and helium are summarized in Table \ref{tab:TabThermalConductivities}. We took the boundary resistance for borosilicate/helium to be the same as for quartz/helium since it depends primarily on the acoustic impedance mismatch. For the superfluid in the channel, the thermal resistance depends on the thickness of the channel. For a cylindrical channel of diameter $d$, it can be approximated \cite{Greywall1981} by $\kappa= 20\,d\,T^3$. This is applied to our channel geometry by taking $d\approx h_c$.

	For both the quartz and borosilicate resonators, the thermal resistance through the helium channel is much larger than the resistance through the substrate, and Eq.~\ref{eq:TotalThermalResistance} can be reduced to ${R_{th}} = {R_k + R_b/2}$.  In the case of the quartz substrate, since the Kapitza resistance $R_{k}$ is almost two orders of magnitude larger than the thermal resistance of the bulk $R_{b}$,  Eq.~\ref{eq:TotalThermalResistance} can be further reduced to $R_{th} = R_{k}$.  In contrast, the Kapitza resistance of borosilicate glass is much smaller than its bulk thermal resistance and the total thermal resistance can be reduced to $R_{th} = R_{b}/2$.

\section*{Supplement F: Dissipation in the borosilicate substrate}
\begin{figure}[b]
			\includegraphics[width=\textwidth]{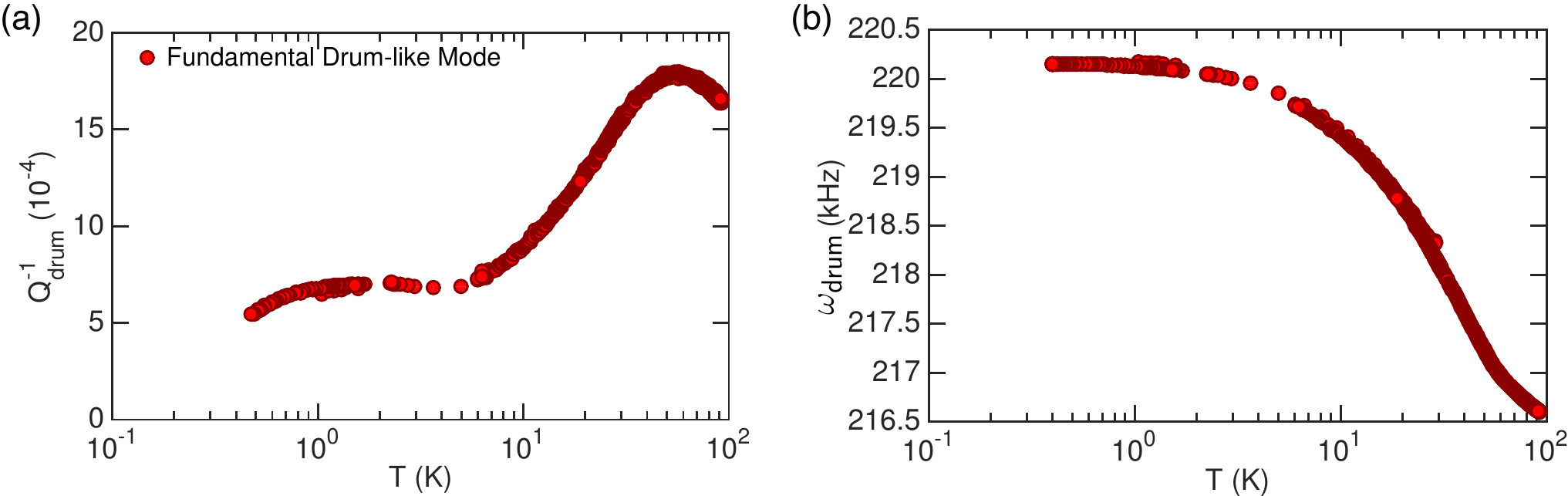}
			\caption{\linespread{1.0} With the borosilicate device evacuated of liquid helium, the mechanical resonance of the fundamental drum-like mode was measured as a function of temperature. (a) The dissipation, $Q_{drum}^{-1}$, of the drum-like mode as a function of temperature. (b) Resonance frequency, $\Omega_{{drum}}$, of the drum-like mode as a function of temperature. As explained in the text, the features of those two curves can be explained by the standard tunneling model of two-level systems \cite{Classen1994}.}\label{fig:DrumLike}
	\end{figure}

	The quality factor of the Helmholtz resonance for the borosilicate device plateaus at a value of $Q=1850$, below 1~K. To understand this limitation, we studied the behavior of the fundamental drum-like mode of an empty device, Fig.~\ref{fig:DrumLike}. In this situation, the damping arises from internal dissipation in the substrate. The resonance frequency, $\Omega_{{drum}}$, of this mechanical mode is approximately $220$~kHz at the base temperature of our $^3$He fridge ($T=475$~mK). Up to $\simeq 5$~K, $\Omega_{{drum}}$ shows a weak temperature dependence as it changes by less than 0.2\%. Above 5~K and up to 100~K, $\Omega_{{drum}}$ only decreases by 2\%.

	The dissipation, $Q_{drum}^{-1}$, of the drum-like resonance displays a more complex temperature dependence, and yet the behavior of both the $Q_{{drum}}^{-1}$ and $\Omega_{{drum}}$ can be understood through the standard tunneling model of two-level systems for amorphous materials \cite{Classen1994}. Here, the analogy with the measurements of Ref.~\cite{Classen1994} is striking, as $Q_{{drum}}^{-1}$ reproduces all the standard features of such systems. From $T=475$ mK to 1 K the dissipation increases with increasing temperature. Above 1 K and up to 5 K, $Q_{{drum}}^{-1}$ plateaus at a value of $7\times 10^{-4}$. Finally a large peak is observed with a maximum value of $18 \times 10^{-4}$ at 55 K.

	It is important to notice that the drum-like mode has a resonance frequency $\approx50$ times higher than the Helmholtz resonance frequency, and the frequency dependence of the dissipation must be considered. The dissipation of two-level systems as a function of frequency has been measured by several groups \cite{Classen2000,Fefferman2008} in amorphous SiO$_2$. Their measurements show that as the frequency is lowered, the temperature of the transition into the plateau region is lowered, down to $100$ mK at $5$~kHz. The value of the dissipation in the plateau region, however, remains unchanged.  Therefore, we expect that in the temperature range (from 475 mK to $T_{\lambda}$), where the Helmholtz resonance is observed, the internal dissipation of the borosilicate is equal to that in the plateau region, \emph{i.e.} $7\times 10^{-4}$. In supplement C we showed that $1/(1+\Sigma)=77$\% of the potential energy is stored in the substrate, therefore we predict that the Helmholtz quality factor in the borosilicate device should be $Q_{int} = (1+\Sigma)\times Q_{{drum}} = (0.77 \times 7\times 10^{-4})^{-1} = 1850 $. This is in good agreement with the value of $1800 \pm 300$ measured, confirming that the limitation in the quality factor of the resonator is indeed coming from the internal dissipation of the borosilicate.

\section*{Supplement G: Dissipation in the quartz substrate}
	\subsection{Two-level systems}

	The Helmholtz resonance in the borosilicate device was found to be limited by internal dissipation in the substrate material due to two-level systems. Fabricating a Helmholtz resonator from single crystal Z-cut quartz substantially reduced the losses of the resonator, however we found that the low temperature behavior is still dominated by internal dissipation in the substrate.  Modeling of the low temperature dissipation is based on a two-level systems ensemble model, similar to the one used in Ref.~\cite{Tao2014} to describe dissipation in single crystal diamond nanoresonators. When the energy splitting of the two-level systems, $\Delta E$, is modulated by the mechanical oscillations, energy is transferred to the two-level systems, with an efficiency depending on the occupation probability of the two-level systems. The quality factor, $Q_{int}$, associated with this process is given by:
	\begin{equation}
		Q^{-1}_{int} = \frac{1}{(1+\Sigma)}\times Q^{-1}_A \times\frac{ \exp{\left(-\frac{T_0}{T}\right)} }{ \exp{\left(\frac{T_0}{T}\right)} + \exp{\left(-\frac{T_0}{T}\right)} }.
	\end{equation}
\noindent 
Here $T_0 = \Delta E/k_B$ with $k_B$ the Boltzmann constant, $Q^{-1}_A$ is the maximum dissipation in the substrate material such that $Q^{-1}_{int} (T=\infty) = Q^{-1}_A / 2$, and $(1+\Sigma)$ is the percentage of potential energy stored in the bending of the substrate, as explained in supplement C. Applying this model to the low temperature portion ($<500$~mK) of the dissipation of the Helmholtz resonance in the quartz device yields $Q_A=1.5\times10^4$ and $\Delta E = 1.3$~GHz.
	
	\subsection{Drum-like mode dissipation}
	
	\begin{figure}[h]
		\includegraphics[width=\textwidth]{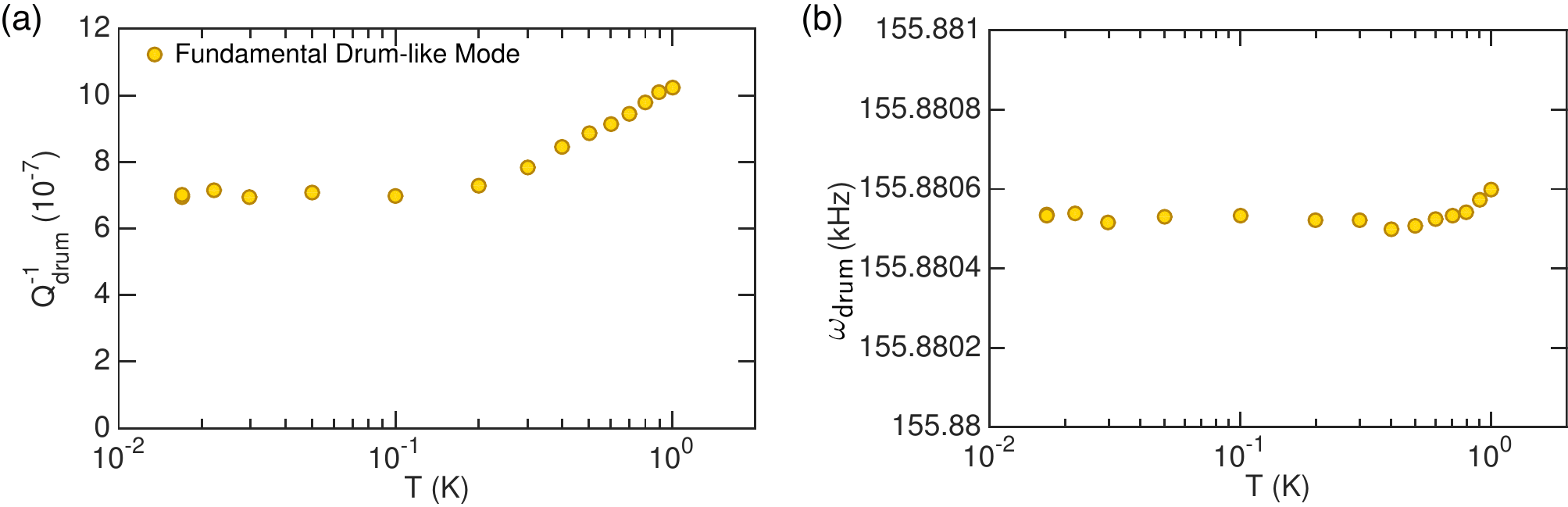}
		\caption{\linespread{1.0} The mechanical resonance of the fundamental drum-like mode for the quartz resonator. (a) Dissipation, $Q^{-1}_{{drum}}$, of the drum-like mode as a function of temperature. (b) Resonance frequency, $\Omega_{{drum}}$, of the drum-like mode as a function of temperature.}\label{fig:DrumLikeQuartz}
	\end{figure}
	Measurement of the drum-like mode, similar to the one performed for the borosilicate device, was made using the quartz device, Fig.~\ref{fig:DrumLikeQuartz}. Due to the thinner quartz substrate, the resonance frequency is lower at $\Omega_{{drum}}=155$~kHz, Fig.~\ref{fig:DrumLikeQuartz}(a). It should be noted that the temperature range ($15$~mK to $1$~K) covered by this measurement is different than the temperature range used in the case of the borosilicate resonator ($450$~mK to $100$~K) due to the cryogenic apparatus used to perform the experiment.

	The internal dissipation of the quartz substrate for the drum-like mode measured in Fig.~\ref{fig:DrumLikeQuartz}(b) plateaus at a value of $Q_{{drum}}^{-1} = 7\times10^{-7}$, markedly lower than the dissipation measured for the Helmholtz resonance mode in the quartz device at $3$~kHz. This could be explained by the fact that the dissipation associated with two-level systems in crystalline materials depends on the measurement frequency, as has been observed in single crystal silicon samples \cite{Kleiman1987}. Those measurements showed that as the frequency is increased, the total dissipation is reduced.  Hence, fabricating a resonator with a scaled down geometry would present a twofold advantage. By increasing the Helmholtz resonance frequency, the internal dissipation of the material would be reduced. Also the contribution of the substrate to the motion would also reduced, decreasing the effective damping coming from the material.

\end{document}